\documentclass[prb,twocolumn,showpacs]{revtex4}

\usepackage[utf8]{inputenc}
\usepackage[T1]{fontenc}

\usepackage{multirow}	
\usepackage{enumerate}	
\usepackage{Tabbing}	
\usepackage{relsize}	
\usepackage{textcomp}	

\usepackage{amsmath}
\usepackage{amssymb}
\usepackage{bm}		

\usepackage{hyperref}
\usepackage{url}

\usepackage{dcolumn}	
\usepackage{longtable}	
\usepackage{booktabs}	

\usepackage[pdftex]{graphicx}
\usepackage{rotating}	
\usepackage{float}	

\usepackage{color}

\newcommand{\average}[1]{\ensuremath{\left\langle#1\right\rangle}}

\newcommand{\matrixel}[3]{\ensuremath{\left\langle #1 \vphantom{#2#3} \right| #2 \left| #3 \vphantom{#1#2} \right\rangle}}

\newcommand{\mc}[3]{\multicolumn{#1}{#2}{#3}}

\renewcommand{\vec}[1]{\ensuremath{\mathbf{#1}}}

\newcommand{\size}{0.78\linewidth}

\begin{document}

\title{Metallization of atomic solid hydrogen within the~extended Hubbard model with renormalized Wannier wave functions}
\author{Andrzej P. K\k{a}dzielawa}
\email{kadzielawa@th.if.uj.edu.pl}
\affiliation{                                                                 
Marian Smoluchowski Institute of~Physics, Jagiellonian University, ul. Reymonta 4, PL30059 Krak\'{o}w, Poland
}

\date{\today}

\begin{abstract}
We refer to our recent calculations (Eur. Phys. J. B, \textbf{86}, 252 (2013)) of metallization pressure of the~three-dimensional
simple-cubic crystal of atomic hydrogen and study the~effect on the~crucial results concocting from approximating the~$1s$ Slater-type orbital function with
a~series of $p$ Gaussians. As a result, we find the~critical metallization pressure $p_C = 102\ GPa$. The latter part is a discussion of the~influence
of zero-point motion on the~stabilizing pressure. We show that in our model the~estimate magnitude of zero-point motion carries a little effect
on the critical metallization pressure at zero temperature.
\end{abstract}


\maketitle

\section{Motivation}\noindent
This year we are celebrating the~$50$th anniversary of \text{the~Hubbard model}, a second-quantization language to describe strongly correlated
systems provided independently by Hubbard \cite{Hubbard1}, Gutzwiller \cite{Gutzwiller1, Gutzwiller2} and Kanamori \cite{Kanamori}. This description
shed some light on many-body quantum systems, in particular on the~localization--delocalization transitions of fermions states
in the~solid-state \cite{MITorg,MIT,Gebhard,Imada}, and optical-lattice \cite{Bloch} systems. This transition is called
the~\textit{Mott} or  \textit{Mott-Hubbard} transition.

In the~series of papers \cite{Kurzyk,Spalek,Kadzielawa}, we have conducted model calculations combining both the~Mott \cite{MITorg} and the~Hubbard \cite{Hubbard2}
aspects of the~phase transition, within the~extended Hubbard model, with a~simultaneous renormalization of the~single-particle Wannier basis,
connecting first- and second-quantization approach. In \cite{Kadzielawa} we obtained, using proposed model, the~critical metallization pressure
$p_C = 97.7\ GPa$ required to stabilize the~atomic-hydrogen-like crystal, while having both the~Mott ($n_C ^{1/3} a_B \approx 0.2$) and the~Hubbard 
($U \approx W$) criteria satisfied at the same time. Thus, those two criteria represent two sides of the same coin.

Ever since Ashcroft proposed an explanation for grater-than-expected magnetic field of Jovian planets \cite{Ashcroft} by applying the~BCS theory to
the~metallic hydrogen, the~pursuit of the~metallization of this element began. Predicted by Wigner and Huntington in 1935 \cite{WignerHydrogen}
the~conducting phase of hydrogen is claimed to have various properties, including hypothesis of being superconducting up to the~room temperature \cite{Ashcroft}. 

In this paper we briefly describe the~model in Section \ref{sec:model}. Then in Section \ref{sec:gaussians} we review the~validity of approximations
made in \cite{Kadzielawa} and show that they were in fact sufficient (explicitly redoing all calculations and showing no qualitative changes).
We also show that both Mott and Hubbard criteria of localization-delocalization transition are satisfied.
In Section \ref{sec:ZPM} we estimate the~magnitude of zero-point motion energy, omitted in our calculations to test the~strength of our results,
keeping in mind the~possibility of quantum melting of hydrogen.

\section{Model}\noindent
\label{sec:model}
  We start with the~extended Hubbard Hamiltonian describing a single-band hydrogen system \cite{Kurzyk,Spalek,Kadzielawa}:
  \begin{equation}
   \label{eq:HubbHam}
\begin{split}
    \mathcal{H} = &\epsilon_{a} \sum_i n_i + \sum_{i\neq j , \sigma} t_{ij} a^{\dagger}_{i \sigma} a^{}_{j \sigma} + U \sum_i n_{i\uparrow} n_{i\downarrow}\\
		  &+ \sum_{i < j} K_{ij} n_i n_j + \sum_{i < j} \frac{2}{R_{ij}},
\end{split}
  \end{equation}
where $t_{ij}$ is the~hopping integral, $U$ the~intraatomic interaction magnitude, $\epsilon_a$ the~atomic energy per site, and
${2}/{R_{ij}}= {2}{\left| \vec{R}_j - \vec{R}_i \right|^{-1}}$ ion-ion interaction corresponding to the~classical Coulomb repulsion (in atomic units).

We have the~total number of electrons $N_e = \sum_i n_i$, and define the~deviation from one-electron-per-atom configuration
$\delta n_i = n_i - 1$. We rearrange \cite{RycerzPhD}
\begin{equation}
 \label{eq:rearK}
  \begin{split}
    \sum\limits_{i < j} {K_{ij} n_i n_j } = &\sum\limits_{i < j} {K_{ij} } \delta n_i \delta n_j  + N_e \frac{1}{N}\sum\limits_{i < j} {K_{ij} } \\
  + &(N_e  - N)\frac{1}{N}\sum\limits_{i < j} {K_{ij} }.
\end{split}
\end{equation}

For half band-filling $n= N_e/N = 1$ the~latter part disappears, and we can write $\sum_{i < j} {K_{ij}} \approx \sum_{i < j} {K_{ij} n_i n_j}$,
thus introducing the~\emph{effective} atomic energy per site
$\epsilon_a^{eff} = \epsilon _a  + \frac{1}{N}\sum\limits_{i < j} \left( K_{ij}  +\frac{2}{R_{ij}} \right)$. Let us rewrite
 the~Hamiltonian \eqref{eq:HubbHam} is a following manner
  \begin{equation}
   \label{eq:ExtHubb}
\begin{split}
    \mathcal{H} = &\epsilon^{eff}_{a} \sum_i n_i + \sum_{i\neq j , \sigma} t_{ij} a^{\dagger}_{i \sigma} a^{}_{j \sigma} + U \sum_i n_{i\uparrow} n_{i\downarrow}\\
		  &+ \frac{1}{2} \sum_{i \neq j} K_{ij} \delta n_i \delta n_j.
\end{split}
  \end{equation}
Since we are interested in calculating explicitly the~average value $\average{\mathcal{H}}$, we note that close to the~metal--insulator boundary
$\average{\delta n_i \delta n_j} \approx 0$, hence we disregard this term in the~calculation of energy.

\subsection{Wave-Function Optimization}\noindent
\label{sec:opt}
To calculate the~microscopic parameters $\epsilon_a$, $t_{ij}$, $K_{ij}$, $U$ of the~Hamiltonian \eqref{eq:ExtHubb} we choose
the~basis of the~orthogonalized-to-the-nearest-neighbors Wannier $w_i$ functions constructed from $1s$ Slater-type orbitals (STO) $\Psi_i$
  \begin{align}
   \label{eq:wannier}
    w_{i} \left( \vec{r} \right) = \beta \Psi_{i} \left( \vec{r} \right) - \gamma \sum_{j=1}^{z} \Psi_{j} \left( \vec{r} \right),
  \end{align}
where $\beta$ and $\gamma$ (see \cite{Kurzyk} eqs. $(24)$ and $(25)$) are mixing parameters specified for the~topology of the~crystal, and depending
explicitly on the~overlap integrals of the~single-particle functions. $z$ is the~number of nearest neighbors.

\begin{figure}
\includegraphics[width=\size]{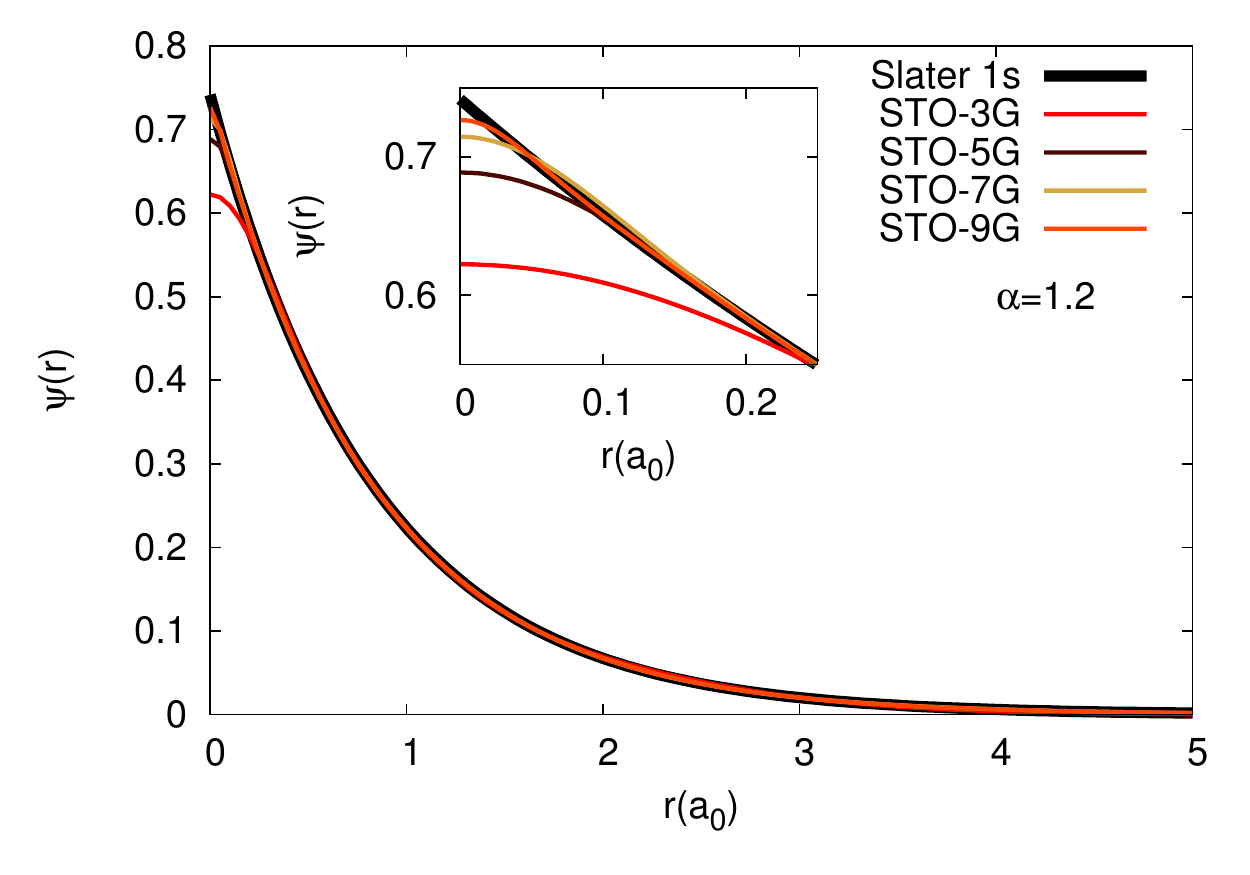}
\caption{Approximations of Slater $1s$ function centered on-site with different Gaussian resolution $p$ (see \eqref{eq:gaussian} and
Tab.~\ref{tab:01}) for $\alpha=1.2$ with respect to distance $r$ from the~ion. Inset: details for small distances. Note that
the~biggest contribution to the~error is given by the~part close to the~node, hence small total error after integrating over whole space.
As expected the~$9$ Gaussian basis (STO-$9$G) is far the~best approximation.}
\label{fig:slater}
\end{figure}

Obtaining the~microscopic parameters from the~first principles requires several integrations, since
\begin{subequations}
 \label{eq:defs}
 \begin{align}
 \label{eq:def_eps}
  \epsilon_a &= \matrixel{w_i}{H_1}{w_i},\\
 \label{eq:def_t}
  t_{ij} &= \matrixel{w_i}{H_1}{w_j},\\
 \label{eq:def_U}
  U &= \matrixel{w_i w_i}{{2}{\left| \vec{r}_1 - \vec{r}_2 \right|^{-1} }}{w_i w_i}, \\
 \label{eq:def_K}
  K_{ij} &= \matrixel{w_i w_j}{{2}{\left| \vec{r}_1 - \vec{r}_2 \right|^{-1} }}{w_i w_j},
\end{align}
\end{subequations}
where $H_1$ is the~Hamiltonian for a single particle in the~system, and ${2}{\left| \vec{r}_1 - \vec{r}_2 \right|^{-1} }$
interparticle interaction in atomic units. Calculating \eqref{eq:defs} with basis as given in \eqref{eq:wannier} requires solving
very complicated series of integrals and can be simplified by approximating STO with a series of Gaussian functions
  \begin{align}
   \label{eq:gaussian}
    \Psi_{i} \left(  \vec{r} \right) = &\sqrt{\frac{\alpha^3}{\pi}} e^{- \alpha \left| \vec{r} - \vec{R_i} \right| } \approx \notag\\
	      &\alpha ^{\frac{3}{2}} \sum_{a=1}^{p} B_a \left( \frac{2 \Gamma_a ^2}{\pi} \right) ^{\frac{3}{4}} e^{- \alpha^2 \Gamma_a^2 \left| \vec{r} - \vec{R_i} \right|^2},
  \end{align}
 where $B_a$ and $\Gamma_a$ are parameters found by minimizing energy of the~single atom ($\mathcal{H}_1 \overset{a.u.}{=} -\bigtriangledown ^2 - {2}{\left| \vec{r} - \vec{R_i} \right| ^{-1}}$).
$p$ is a number of Gaussian functions used for the~approximation.
$\alpha$ is the~inverse function size and will remain a variational parameter, allowing us to renormalize the~ground state function
to find the~minimal energy for given lattice parameter $R$. For the~sake of completeness we explicitly illustrate the~quality of the-approximation (Figure~\ref{fig:slater})
and the coefficient for different STO-$p$G basis (Table~\ref{tab:01}).

\begingroup
\squeezetable
\begin{table*}
 \caption{\label{tab:01} $B_a$ and $\Gamma_a$ coefficient obtained by minimizing the~single-particle energy with wavefunctions given by \eqref{eq:gaussian}.}

\begin{tabular}{|c||c|c||c|c||c|c||c|c|}
\hline
\mc{2}{|c||}{STO-$3$G} & \mc{2}{c||}{STO-$5$G} & \mc{2}{c||}{STO-$7$G} & \mc{2}{c}{STO-$9$G} & \\\hline
\mc{1}{|c|}{$B_a$} & \mc{1}{c||}{$\Gamma_a^2$} & \mc{1}{c|}{$B_a$} & \mc{1}{c||}{$\Gamma_a^2$} & \mc{1}{c|}{$B_a$} & \mc{1}{c||}{$\Gamma_a^2$} & \mc{1}{c|}{$B_a$} & \mc{1}{c}{$\Gamma_a^2$} & \mc{1}{||c|}{a} \\\hline\hline
\mc{1}{|c|}{0.7079069} & \mc{1}{c||}{0.4037496} & \mc{1}{c|}{0.4862397} & \mc{1}{c||}{0.3428813} & \mc{1}{c|}{0.3347926} & \mc{1}{c||}{0.3073439} & \mc{1}{c|}{0.2333815} & \mc{1}{r}{0.2832535} & \mc{1}{||c|}{1} \\\hline
\mc{1}{|c|}{0.3460096} & \mc{1}{c||}{0.8919739} & \mc{1}{c|}{0.4687430} & \mc{1}{c||}{0.6489746} & \mc{1}{c|}{0.4947580} & \mc{1}{c||}{0.5341995} & \mc{1}{c|}{0.4735227} & \mc{1}{r}{0.4656983} & \mc{1}{||c|}{2} \\\hline
\mc{1}{|c|}{0.0691531} & \mc{1}{c||}{1.9705714} & \mc{1}{c|}{0.1446282} & \mc{1}{c||}{1.2283203} & \mc{1}{c|}{0.2218991} & \mc{1}{c||}{0.9285009} & \mc{1}{c|}{0.2825582} & \mc{1}{r}{0.7656564} & \mc{1}{||c|}{3} \\\hline
\mc{1}{c}{$ $} & \mc{1}{c|}{$ $} & \mc{1}{c|}{0.0307340} & \mc{1}{c||}{2.3248533} & \mc{1}{c|}{0.0674427} & \mc{1}{c||}{1.6138428} & \mc{1}{c|}{0.1065788} & \mc{1}{r}{1.2588187} & \mc{1}{||c|}{4} \\\cline{3-9}
\mc{1}{c}{$ $} & \mc{1}{c|}{$ $} & \mc{1}{c|}{0.0093803} & \mc{1}{c||}{4.4002717} & \mc{1}{c|}{0.0188009} & \mc{1}{c||}{2.8050467} & \mc{1}{c|}{0.0341750} & \mc{1}{r}{2.0696289} & \mc{1}{||c|}{5} \\\cline{3-9}
\mc{1}{c}{$ $} & \mc{1}{c}{$ $} & \mc{1}{c}{$ $} & \mc{1}{c|}{$ $} & \mc{1}{c|}{0.0038829} & \mc{1}{c||}{4.8754978} & \mc{1}{c|}{0.0099417} & \mc{1}{r}{3.4026852} & \mc{1}{||c|}{6} \\\cline{5-9}
\mc{1}{c}{$ $} & \mc{1}{c}{$ $} & \mc{1}{c}{$ $} & \mc{1}{c|}{$ $} & \mc{1}{c|}{0.0018480} & \mc{1}{c||}{8.4741829} & \mc{1}{c|}{0.0032307} & \mc{1}{r}{5.5943683} & \mc{1}{||c|}{7} \\\cline{5-9}
\mc{1}{c}{$ $} & \mc{1}{c}{$ $} & \mc{1}{c}{$ $} & \mc{1}{c}{$ $} & \mc{1}{c}{$ $} & \mc{1}{c|}{$ $} & \mc{1}{c|}{0.0006094} & \mc{1}{r}{9.1977233} & \mc{1}{||c|}{8} \\\cline{7-9}
\mc{1}{c}{$ $} & \mc{1}{c}{$ $} & \mc{1}{c}{$ $} & \mc{1}{c}{$ $} & \mc{1}{c}{$ $} & \mc{1}{c|}{$ $} & \mc{1}{c|}{0.0004466} & \mc{1}{r}{15.1220138} & \mc{1}{||c|}{9} \\\cline{7-9}
\end{tabular}
\end{table*}
\endgroup

\subsection{Ground-state energy}\noindent
As stated earlier we would like to determine the~inverse wave function size $\alpha$ minimizing the~ground-state energy. To obtain the~values for given
$\alpha$ and the fixed lattice parameter $R$ we use \textit{Statistically-consistent Gutzwiller approximation} (SGA) \cite{SGA}. We
extend the~Gutzwiller approximation Hamiltonian
\begin{equation}
 \label{eq:GAHam}
  \mathcal{H}_{GA} =  \epsilon^{eff}_{a} \sum_{i \sigma} n_{i \sigma} + \sum_{i j \sigma} t_{ij} q_\sigma a^{\dagger}_{i \sigma} a^{}_{j \sigma} + N U d^2 ,
\end{equation}
where the~double occupancy number $d^2 = \average{n_{i\uparrow} n_{i \downarrow}}$  and
\scalebox{0.95}{$q_\sigma = {2 \left(d \sqrt{1-2 d^2-m}+\sqrt{d^2 \left(1-2 d^2+m\right)}\right)^2}/{(1 - m^2)}$}
for $n=1$, by introducing the~Lagrange-multiplier constrains
\begin{equation}
 \label{eq:Lagrange}
 \mathcal{C}_\lambda = -\lambda_m \sum_i \left( m_i - m \right) - \lambda_n  \sum_i \left( n_i - n\right).
\end{equation}
where $m_i \equiv n_{i\uparrow} - n_{i\downarrow}$, $m\equiv \average{m_i}$, $n_i \equiv n_{i\uparrow} + n_{i\downarrow}$, and $n \equiv \average{n_i}$.

Finally, we use the~operator $\mathcal{K} = \mathcal{H}_{GA} + \mathcal{C}_\lambda$ as our effective Hamiltonian. Mean fields $d^2$ and $m$, as well as
the~Lagrange multipliers $\lambda_m$ and  $\lambda_n$, and the~chemical potential $\mu$ are all determined variationally.

Once the~ground-state energy is found as a minimal value for some $\alpha_{min}$, we get the~set of values - the~microscopic parameters
\eqref{eq:defs} in the~ground state. Below we discuss the~properties of our results in comparison to those obtained earlier \cite{Kadzielawa}.

\section{Gaussian basis resolution}\noindent
\label{sec:gaussians}
In our previous approach \cite{Kadzielawa} we favored the~Gaussian basis consisting of $3$ functions. We argued that the~quality of such an approximation 
is sufficient, and that the~numerical effort to obtain results in higher Gaussian resolutions ($p>3$) is unnecessary.
The computational complexity scales
\begin{subequations}
 \label{eq:Complexity}
  \begin{align}
    \label{eq:Complexity2}
    \epsilon_a, t &\propto p^2, \\
    U, K_{ij} &\propto p^4,
    \label{eq:Complexity4}
  \end{align}
\end{subequations}
 where $p$ is the~resolution. Hence the~time of calculating the~full set of data points is increased by a~factor of 200 when replacing
STO-$3$G to STO-$9$G basis.

\subsection{STO-$3$G versus STO-$9$G}\noindent
For our \textit{ab initio} calculations we have selected STO-$9$G basis. It is much better (cf. Figure~\ref{fig:slater}) than STO-$3$G, while time
of the~calculation is still acceptable.
\begin{figure}
\includegraphics[width=\size]{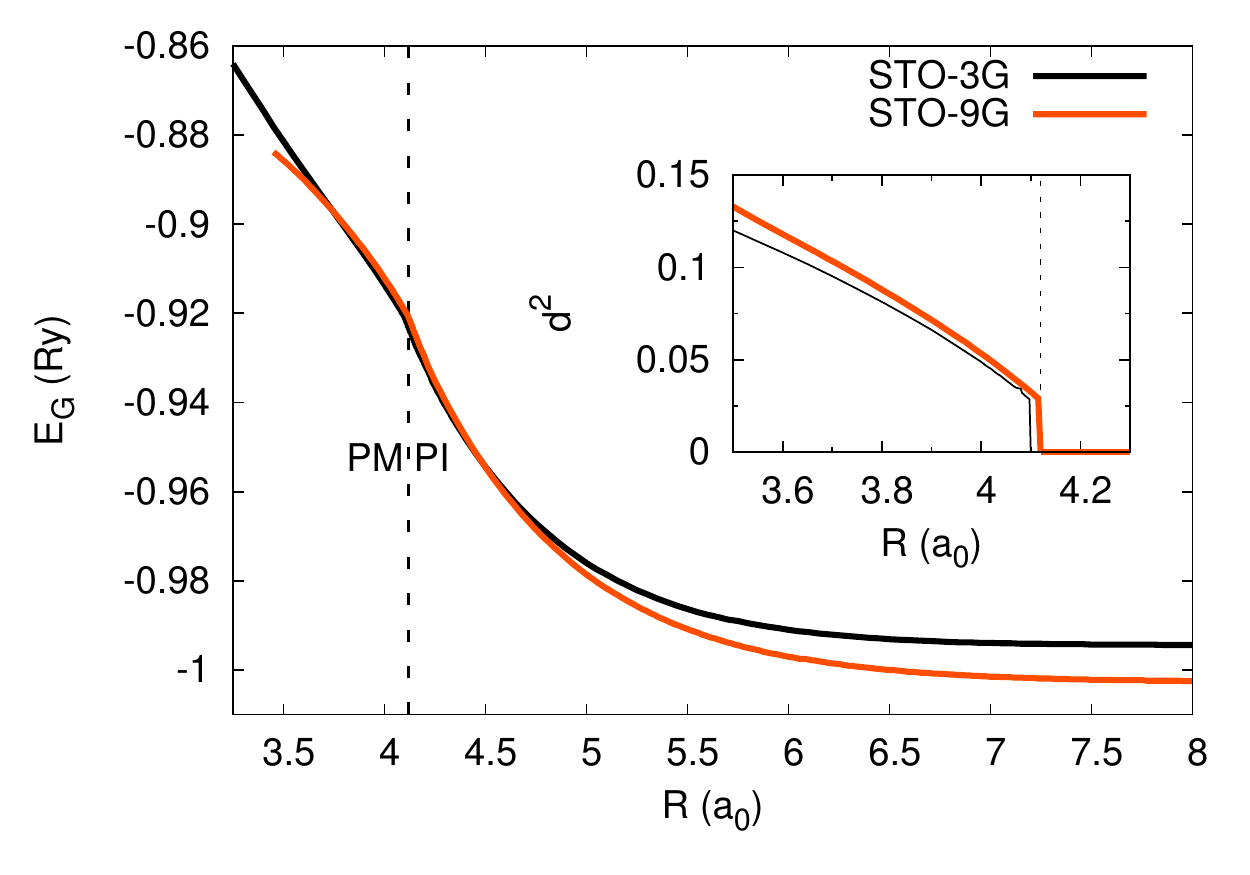}
\caption{Ground-state energy versus lattice parameter $R$ for different STO-$p$G basis. Note more realistic behavior in the~metallic ($R<R_C=4.12 a_0$)
regime with non-trivial $R$ dependance. Inset: Double occupancy mean field versus lattice parameter $R$ for different STO-$p$G basis. Note no qualitative changes of behavior}
\label{fig:EG}
\label{fig:dpow2}
\end{figure}

The dependence of the~ground-state Energy $E_G$ with respect to the~lattice parameter $R$ (Figure~\ref{fig:EG}) is the~main outcome. Similarly to the~previous
case \cite{Kadzielawa}, there are two local minima - one associated with the metallic phase ($d^2 \neq 0$), and one with the Mott insulating phase ($d^2 = 0$). The transition
occurs at $R=R_C=4.12 a_0$ (compared to $R_C^{old}=4.1 a_0$), but its nature is not changed, as it still is a~\textit{weakly} discontinuous transition
(observe the~obvious discontinuity of double occupancy number, cf. inset in Fig.~\ref{fig:dpow2}). 

\begin{figure}
\includegraphics[width=\size]{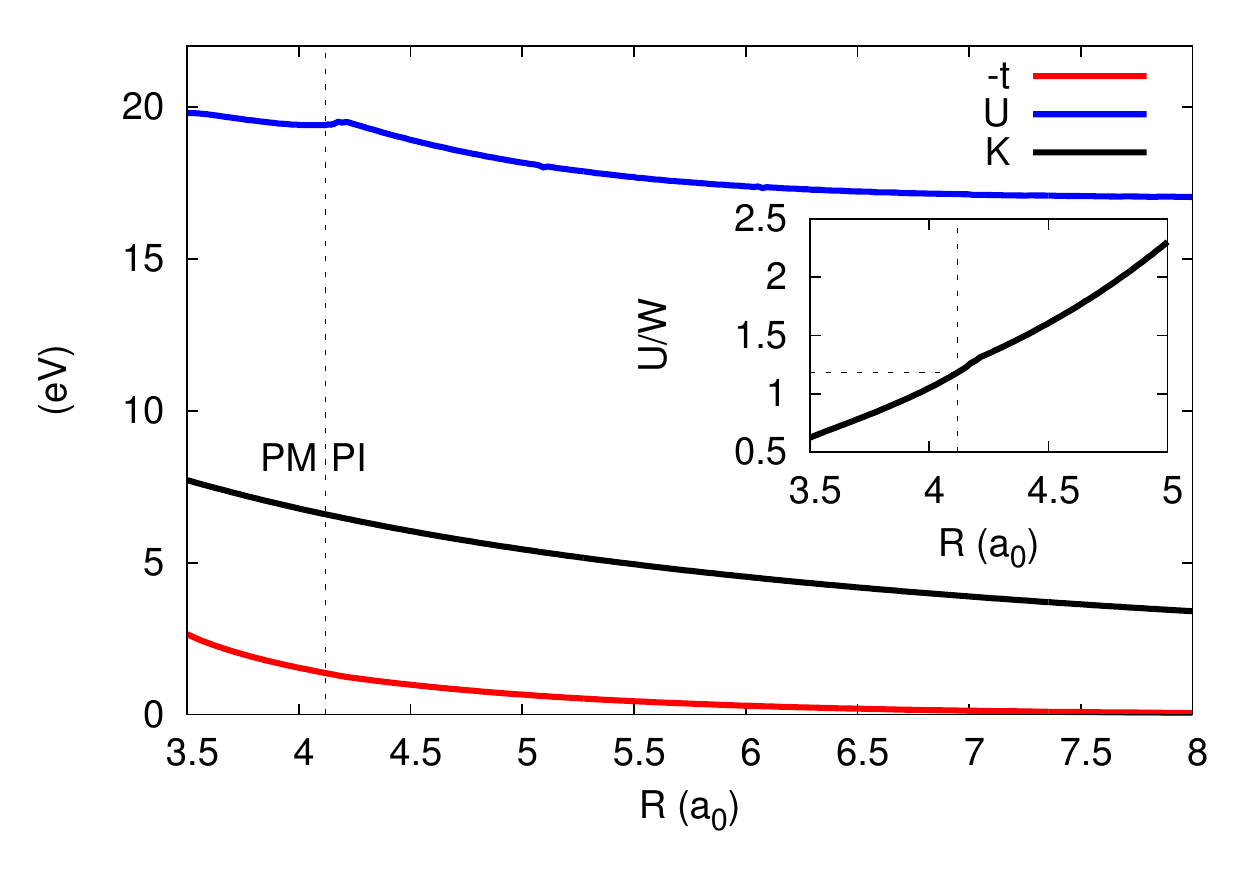}
\caption{The microscopic parameters $t$, $U$ and $K$ versus lattice parameter $R$. Inset: $U/W$ ratio with bandwidth $W=2 z |t|$ and on-site repulsion $U$.}
\label{fig:UtKW}
\end{figure}

In Figure~\ref{fig:UtKW} we plot the~values of the nearest-neighbor hopping $(-t)$, on-site repulsion $U$, and the nearest-neighbor intersite repulsion $K$.
Even though there are no qualitative changes in the~values in comparison with \cite{Kadzielawa} we present this for the~sake of completeness.

In \cite{Kadzielawa} we have shown that our transition satisfies both the~Mott and the~Hubbard criteria for metal--insulator transition.
Below we refer to them while discussing the~new results.

\subsection{The Mott and the Hubbard criteria}\noindent
The original Mott criterion \cite{MITorg,MIT} $n_C^{1/3} a_B \sim 0.2$ can be rewritten by substituting $\alpha^{-1}$ for the~effective Bohr radius
$a_B$ and defining the~particle density as $n_C = R_C ^{-3}$. We get $n_C^{1/3} a_B = R_C ^{-1} \alpha^{-1} \approx 0.22$, a slightly better outcome than
in \cite{Kadzielawa} (as it is predicted with a better accuracy).

As shown in inset to Fig.~\ref{fig:UtKW}, the~ratio $(U/W)$ for critical lattice parameter $R_C=4.12 a_0$ is equal $1.18$ in consistence with \cite{Hubbard2}.

\subsection{Metallization pressure}
\begin{figure}
\includegraphics[width=\size]{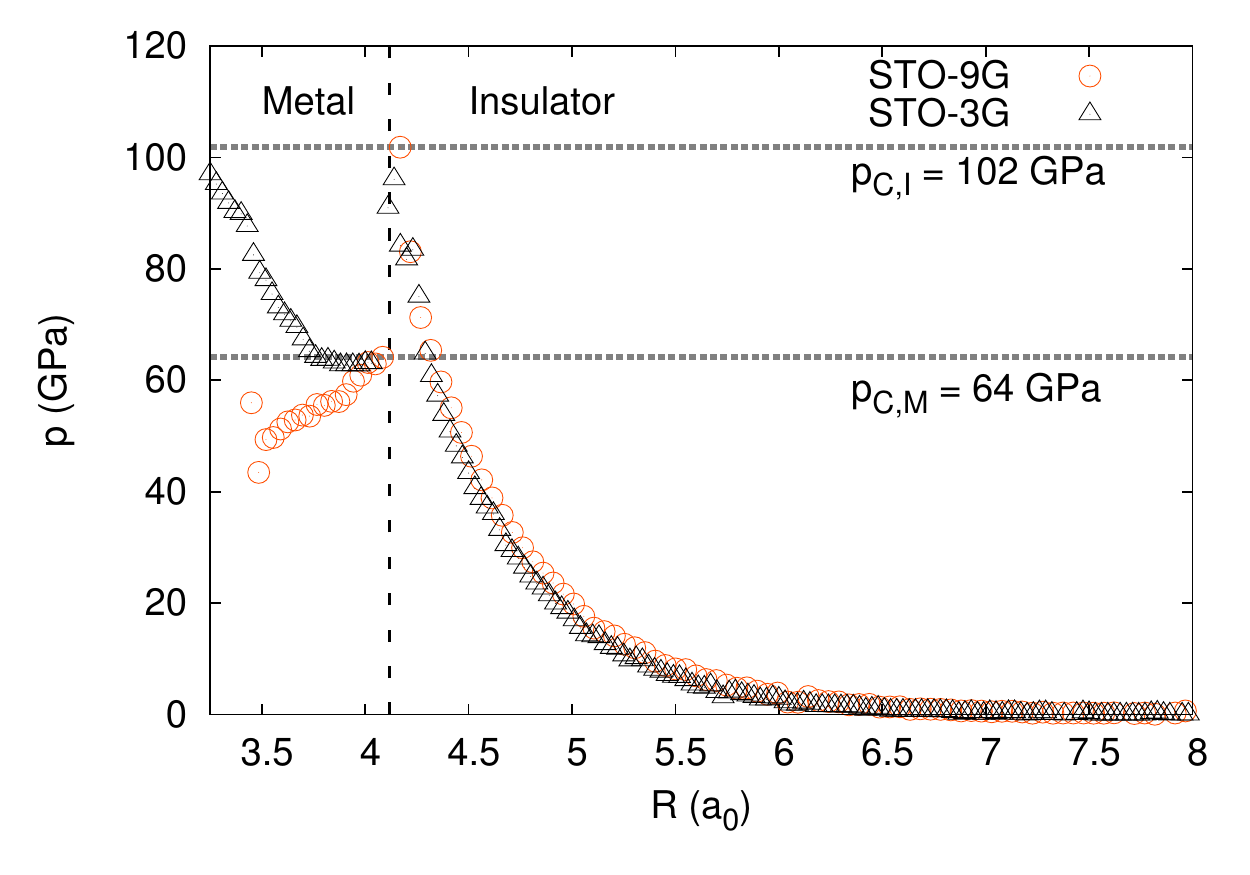}
\caption{Stabilizing pressure for a~simple-cubic atomic solid hydrogen crystal versus lattice parameter $R$ for different STO-\textit{n}G basis. Note only a~slight
change in obtained critical pressure $p_C = 102\ GPa$ for significantly larger STO-$9$G basis. The qualitatively different behavior of stabilizing pressure in the~metallic
($R<R_C=4.12 a_0$) regime is caused by non-trivial behavior of energy in this regime (see Fig. \ref{fig:EG} for details). }
\label{fig:pressure}
\end{figure}

\noindent
Our model represents a~3-dimensional simple-cubic crystal of the~atomic hydrogen (one electron per ion, $1s$ orbitals) undergoing the~Mott--Hubbard transition.
It is clear that the~minimal value of energy (cf. Figure~\ref{fig:EG}) of such a crystal is reached for lattice parameter $R \rightarrow \infty$. Thus one
require external pressure $p$ for its stabilization, that can be obtained classically as the~force per cell $F=\left| - \triangledown_R E_G \right|$
over the~elementary cell area $A/N=R^2$. In Figure~\ref{fig:pressure} we plot such pressure versus lattice parameter $R$ and provide a comparison between
the~previously obtained (STO-$3$G \cite{Kadzielawa}) results and the~new ones.

We have calculated the~metallizing pressure $p_C=102\ GPa$ assuming that our model is static - this assumption is not quite correct within the quantum-mechanical world, where there
is always a non-zero energy of zero-point oscillations. In the~next section we deal with this problem by estimating the~contribution of zero-point motion
to the~total energy.

\section{Zero-point motion energy}\noindent
\label{sec:ZPM}
\begin{figure}
\includegraphics[width=\size]{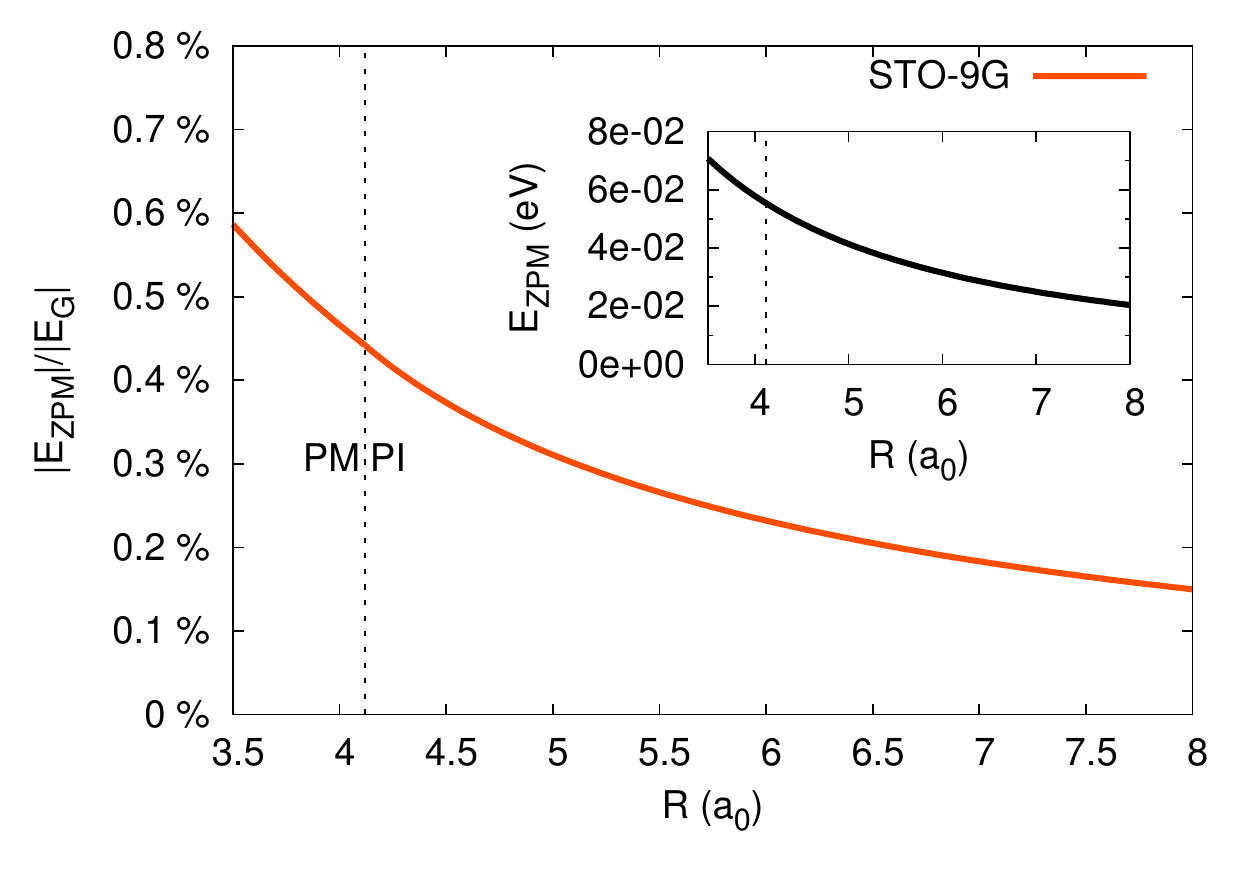}
\caption{The relative magnitude of estimated zero-point motion energy with respect to ground-state energy at given lattice parameter $R$. Note that result below $0.5 \%$ at
the~metal--insulator transition shows that the~correction from ZPM to the~critical pressure can be disregarded. Inset: explicit value of estimated zero-point motion energy.}
\label{fig:EZPM}
\end{figure}
We introduce (following approach similar to \cite{Spalek6}) the~uncertainties of the~momentum $\delta \vec{P}$ and position $\delta \vec{R}$.
The energy of a distortion per ion is

\begin{equation}
 \label{eq:dEnZPM}
  \Delta E =\frac{\delta \vec{P}^2}{2 M_{H^+}} + \frac{1}{2} \sum_\text{\resizebox{0.13\linewidth}{!}{$i \in \{ x , y, z \}$}} \left( \frac{\text{e}^2}{R+\delta R_i} + \frac{\text{e}^2}{R-\delta R_i} \right).
\end{equation}
By applying the~uncertainty relation $\delta \vec{P}^2 \cdot \delta \vec{R}^2 \geq {3 \hbar^2}/{4}$ and minimizing \eqref{eq:dEnZPM} with respect
to $R_i$'s we get a set of local extrema, from which the~global minimum is

\begin{align}
  \label{eq:EZPM}
  \Delta E_0 = 3\frac{\text{e}^2}{R} + \frac{\hbar  \left(4 \sqrt{6} \text{e} M R+\sqrt{M} \sqrt{R} \hbar \right)}{8 M^{3/2} R^{5/2}},
\end{align}
\begin{align}
  \label{eq:RZPM}
  \left| \delta \vec{R}_0 \right| = \sqrt{\frac{3 R^2 }{2 \sqrt{6} \frac{\text{e}}{\hbar} \sqrt{M} \sqrt{R}+ 1}},
\end{align}
where $R$ is the~lattice parameter. The first term of \eqref{eq:EZPM} is related to the~Coulomb repulsion of ions and the~second
$E_{ZPM} \equiv \Delta E_0 - 3{\text{e}^2}{R^{-1}}$ is the~zero-point oscillation energy.

In Figure~\ref{fig:EZPM} we show the~ratio of $|E_{ZPM}|$ to the~ground-state energy $|E_G|$. Since it is slowly-changing and is about two orders of magnitude
smaller than the~ground-state energy, our approach of omitting it in the~calculation of metallization pressure holds.

\section{Conclusions}\noindent
In this paper we established that the~choice of the~STO-$3$G basis in \cite{Kadzielawa} was not influencing results qualitatively, and that the~computational
simplicity and total CPU time conservation are allowing us to examine also a full picture with an external magnetic field, preserving main properties
of the~system. Better accuracy (Subsection~\ref{sec:opt}) increases the~quality of the~results (cf. Figure~\ref{fig:EG}), but does not
change our understanding of the~metal--insulator transition in this model.

The analysis of zero-point motion carried out in Section~\ref{sec:ZPM} reinforces our previous results and suggests that the~energy of oscillations does not
increase the~stabilization pressure significantly.

\section{Acknowledgments}\noindent
I would like to thank Professor J\'{o}zef Spa\l{}ek for critical reading of this paper as well as Dr. Andrzej Biborski and Marcin Abram for 
discussions.

The work was realized as a part of the~TEAM project awarded to our group by the~Foundation for Polish Science (FNP) for the~years 2011-2014.
{
\bibliography{bibliography}
}
\end{document}